\documentclass[10pt]{article}

\usepackage{amsmath}
\usepackage{amssymb}

\usepackage{graphicx}

\usepackage{cite}

\usepackage{color} 

\usepackage[labelfont=bf,labelsep=period,justification=raggedright]{caption}

\makeatletter
\renewcommand{\@biblabel}[1]{\quad#1.}
\makeatother

\date{}

\begin{document}

\begin{flushleft}
{\Large
\textbf{Genome wide identification of regulatory networks associated with 
general cognitive ability using a normalized alignment free similarity measure of promoter regions}

}

Miriam Ruth Kantorovitz$^{1,\ast}$, 
David Tcheng$^{2}$, 
Michael I. Lerman$^{3}$, 
Eric Jakobsson$^{4}$
\\
\bf{1} Department of Mathematics 
University of Illinois, Urbana, IL, USA
\\
\bf{2} National Center for Supercomputing Applications University of Illinois, Urbana, IL, USA
\bf{3} Scientific Board, 
Affina Biotechnologies, Inc. Stamford, CT, USA
\\
\bf{4} Department of Molecular and Integrative Physiology, Department of Biochemistry, UIUC programs in Biophysics, Neuroscience, and Bioengineering, National Center for Supercomputing Applications, and Beckman Institute, University of Illinois, Urbana, IL, USA
\\
$\ast$ E-mail: ruth@math.uiuc.edu
\end{flushleft}

\section*{Abstract}
We show that a normalized alignment free similarity measure, called D2z, can be used to detect potential regulatory relations for gene sets when little is known about the regulatory elements involved. One scenario where such gene sets arise is genome wide association studies (GWAS). In this work 
we consider a gene set from a GWAS on childhood general cognitive ability. We build a co-regulation network for the GWAS genes based on the D2z scores, which shows potential co-regulatory relationships between the genes as well as predict additional genes that are likely to be part of the network. We found that the set of the predicted genes is enriched in genes associated with mental retardation and GO terms such as synapse and neuron development. In particular, we found strong evidence of regulatory connection between the GWAS genes and CHL1, a gene known to be involved in mental retardation. 

\section*{Introduction}

The genetic architecture of childhood general cognitive ability (g) is complex. Many genes with small effects may be involved rather than a few genes with large effects (\cite{Plomin} and references within). In a large genome-wide association study (GWAS), nine SNPs were found to be associated with g~\cite{Plomin}. These loci were mapped to eight genes with non-overlapping promoter regions. In this paper we use a normalized 
alignment free similarity measures, called D2z, applied to promoter regions  of genes,  to predict co-regulation relationship between these genes associated with g. 
 
The similarity of two biological sequences has traditionally been assessed within the well-established framework of alignment. However, when studying gene regulation, one often needs to identify functional relationships between DNA sequences that do not exhibit any statistically significant alignment. This is the case, for example, when comparing cis-regulatory modules (CRMs) in the promoter regions of non-orthologous genes. 
The alignment free similarity measure D2z\cite{KantorovitzISMB, KantorovitzCell} was shown to detect 
 functional similarity between regulatory sequences.
In\cite{KantorovitzISMB} we showed that the D2z method can accurately discriminate functionally related CRMs from unrelated sequence pairs in fruit fly and human. In\cite{KantorovitzCell} the D2z measure was used for genome-wide searches in fruit fly and human to predict  CRMs with a similar function to a given set of CRMs known to mediate a common gene expression pattern, but without the use of known motifs. These predictions  had been validated successfully in vivo.
Various alignment-free similarity measures had been proposed previously (e.g., \cite{Vinga} and references within), however, when applied to real biological data, these methods were unable to discriminate functionally related CRMs from unrelated sequence pairs\cite{KantorovitzISMB}. The main problem with these methods is that they are not ÒnormalizedÓ. That is, the similarity scores are not comparable across sequence pairs drawn from arbitrary background distributions. More recently, other promising normalized alignment free methods have been developed, such as \cite{Reinert, d2zref}.

In this work we test whether the D2z measure, when applied to promoter regions of genes, can be used for detecting possible regulatory relations between genes, without using prior knowledge of the transcription factors (TFs) that may be involved. This is the case for the set of genes implicated in general cognitive ability in the GWAS, where the interactions or co-regulation relationship between the genes is not known. 
As in the CRM case, the idea is that if two genes have similar promoter regions, 
then these promoter regions are likely to contain binding sites for the same (unknown) TFs that target these genes and hence the two genes are likely to be co-regulated. 
However, unlike CRMs, which usually contain multiple binding sites for relevant TFs, a promoter region may not be as enriched in TF binding sites, and therefore the similarity is harder to detect.
We show that our method detects co-regulatory relationship between the eight genes associated with g, as well as connects them to genes that are known to be involved in brain development or intellectual disabilities.

From previous work~\cite{ML1, ML2, ML3}, we were particularly interested  in regulatory connections between the gene Close homologue of L1 (CHL1, also known as CALL) and the cognition  genes from the GWAS study.
CHL1 is a neural recognition molecule that plays important roles in cell migration, axonal growth, and synaptic remodeling. It has been associated with mental retardation, reduced intelligence, schizophrenia, memory and behavior disorders\cite{ML1, ML2, ML3, CALL1, CALL2, CALL3, CALL4, CALL5, CALL6, CALL7}. 
Our method showed a strong connection between CHL1 and the cognition genes. 
We perform a genome-wide search for additional genes that are potentially co-regulated with the genes found in the GWAS. We found that the set of the predicted genes is enriched in genes associated with mental retardation and GO terms such as synapse, exonogenesis and neuron development.


The advantage of our approach is that it does not rely on TF databases and does not use prior knowledge of the TFs that may be involved in the network. Since the only input we use is the genes{Õ} promoter regions, this method can be readily applied to any genome. Other normalized alignment free methods (e.g. \cite{Reinert, d2zref}) could potentially be used in the same manner. This approach, which also does not use information from the coding regions, can be combined with methods that use such information, to better understand the networks in question.


\section*{Results and Discussion}

In the GWAS study\cite{Plomin}, eight genes with non-overlapping promoter regions were associated with general cognitive ability (g), which were also in the DBTSS database: RXRA, CARS, CTNNA3, TMCC3, MAP3K7, STK10, NR2F1 and FERMT1. We first construct the co-regulation graph for these genes as follows.

\subsubsection*{Constructing the co-regulation graph}
 For a given set of genes, we construct a co-regulation graph using the promoter region of the genes (see Methods for details). Briefly,  each promoter region from the dataset ("probe") is scored against the promoter regions of all other human genes in the database DBTSS\cite{dbtss} using the D2z similarity measure~\cite{KantorovitzISMB}. For each probe gene, the genes associated with the top scoring promoter sequences  are retrieved. This relationship is captured by a directed edge in the co-regulation graph, from the probe gene to each of the top scoring genes. Thus, the edges in the co-regulation graph represent a co-regulation relationship. A highly connected graph means that the probe genes have many commonly co-regulated genes and therefore these genes are likely to be part of a regulatory network. In this work we used the top 30 scoring genes for each probe, which is about 0.1 percent of the sequences in the DBTSS database.
 
\begin{figure}
\begin{center}
\includegraphics[width=10.cm,height=15.cm,angle=0,clip]{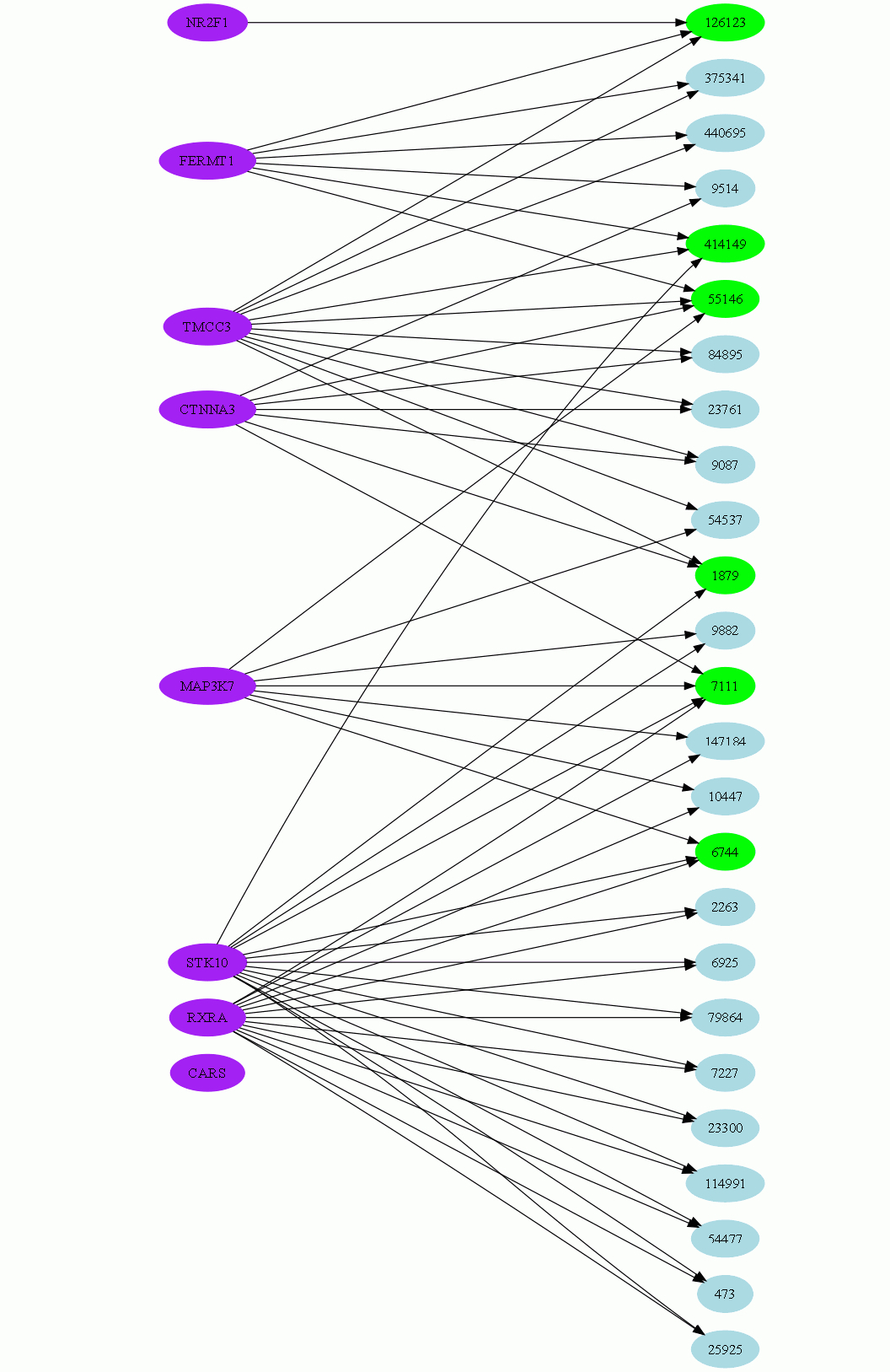}
\caption{Co-regulation graph for the GWAS gene set. The purple nodes in the left column are the probes from the GWAS gene set. The nodes in the right column are genes that are commonly co-regulated by at least 2 probes. The green nodes are genes that are commonly co-regulated by  3 or more probes.
}
\label{fig1}
\end{center}
\end{figure}

Figure~\ref{fig1} shows the co-regulation graphs for these genes. 
We see that 7 of the 8 probes are connected, with 25 shared nodes ($p<10^{-5}$). 
One gene, CARS, is not connected to the rest of the graph. 

\subsubsection*{Relations with CHL1 and genome-wide discoveries}
\begin{figure}
\begin{center}
\includegraphics[width=15.cm,height=20.cm,angle=0,clip]{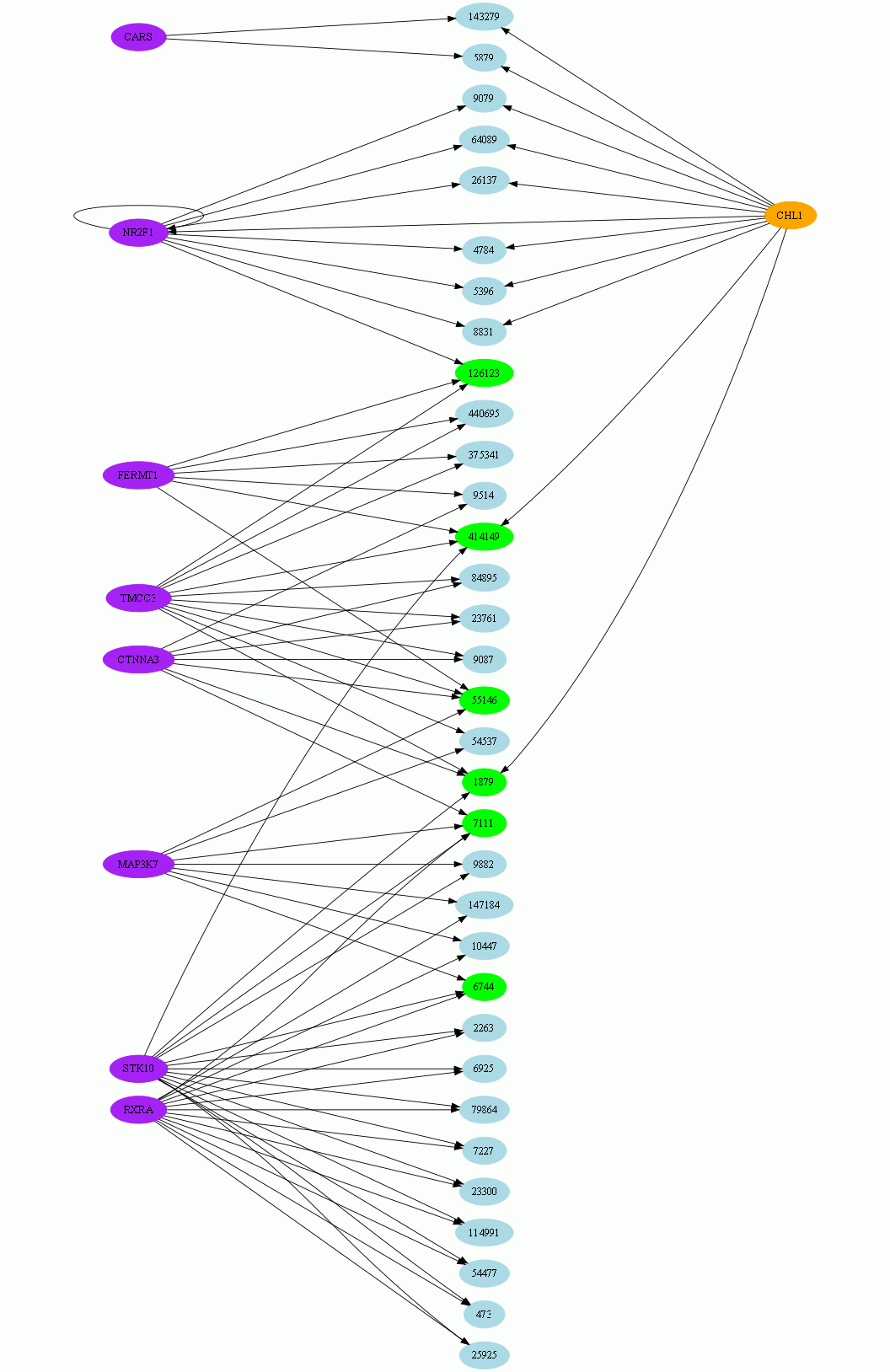}
\caption{Co-regulation graph for the GWAS gene set and CHL1.  The purple nodes in the left column are the probes from the GWAS gene set. The orange node on the right is the probe CHL1.   The nodes in the middle column are genes that are commonly co-regulated by at least 2 probes. The green nodes are genes that are commonly co-regulated by  3 or more probes.
}
\label{fig2}
\end{center}
\end{figure}

To find potential co-regulatory relationship between CHL1, which is a gene known to be involved in mental retardation, schizophrenia, memory, and social behavior\cite{ML1, ML2, ML3, CALL1, CALL2, CALL3, CALL4, CALL5, CALL6, CALL7}, and the GWAS cognition genes, we constructed the co-regulation graph for CHL1 with the eight cognition genes (Figure~\ref{fig2}).
Interestingly, when we added CHL1 to the probe gene set,
the co-regulation graph became connected with 33 shared nodes ($p<10^{-5}$). In addition, CHL1 had a shared node with seven of the eight probes from the GWAS study, and a direct edge to one of the probes:  NR2F1. Among the genes that CHL1 and NR2F1 share in the co-regulation graph (that is, genes that are commonly co-regulated with both, CHL1 and NR2F1) were genes that are known to be involved in learning disabilities and brain development, such as, SYNGAP1\cite{SYNGAP1, SYNGAP1ref2} and NFIX\cite{NFIX}. A complete list of the top 30 scoring genes for each probe is provided in supplementary material.
This list includes genes that are potentially co-regulated with CHL1 but are not in the co-regulation graph since they were not commonly co-regulated with genes from the GWAS data set.  

To discover more genes with potential co-regulatory properties as CHL1 with respect to the GWAS gene set, we searched genom-wide for genes that are commonly co-regulated with at least one of the genes in the GWAS gene set (i.e., genes with a direct edge to the probe set in the co-regulation graph). We found
about 900 such genes, which are potentially part of the regulatory network involved in general cognitive ability (see supplementary material). 

Using the functional annotation tool in the Database for Annotation, Visualization, and Integrated Discovery (DAVID)\cite{DAVID}, we found that this set of genes is enriched in GO terms such as synapse ($p=1.30*10^{-9}$), neuron development ($p=1.60*10^{-5}$), exonogenesis ($p=1.2*10^{-5}$) axon guidance ($p=8.00*10^{-5}$) and cell adhesion ($p=1.00*10^{-4}$). 
In particular, this set of genes is enriched in genes associated with mental retardation in the OMIM database ($p=4.30*10^{-5}$), suggesting that the predicted genes to be co-regulated with the GWAS gene set may indeed be part of a network involved in g. The list of the mental retardation genes contained in our set of predicted genes is given in Table~\ref{Table1}.

\subsection*{Conclusions}
The D2z method is a promising tool for de novo inference of co-regulated networks of genes without reference to prior knowledge of the identity of transcription factor binding sites or transcription factors.  Because of its computational efficiency and its comprehensiveness it should be generally useful as an adjunct to existing methods of transcription network inference, and especially useful for inference of networks from genome wide association studies.  A major weakness to date of genome-wide association studies lies in poor statistics available for inference of underlying genomic bases when phenotypes are dependent on multiple genes, or interactions between genes.  By inferring co-regulation, the D2z method serves effectively to screen, validate and extend faint signals available from genome-wide association studies.  As noted in the text, other alignment-free methods could potentially be used in the same manner.  Comparison of the specific methods for this purpose awaits further research.


\section*{Materials and Methods}
\subsubsection*{Constructing the co-regulation graph}
Given a data set containing a few human genes that are believed to be involved in the network, the promoter region of each gene in the dataset (referred to as a ÒprobeÓ gene) is extracted, using the DBTSS database version 6 \cite{dbtss}  including redundancies (32,122 sequences). We used the default setting in DBTSS for the length of the sequences, which is -1000 to +200. Each of these ÒprobeÓ promoter regions is scored against the promoter regions of all other human genes in DBTSS using the D2z similarity measure\cite{KantorovitzISMB}. The D2z measure is a normalized alignment free similarity measure that is based on the frequencies of k-words in the sequences. The k-words represent potential binding sites of (unknown) TFs. In this analysis, the parameter k was set to be 5, representing the core length of a TFBS\cite{KantorovitzISMB}.
For each gene in the probe set we retrieve the genes associated with the top $n$ scoring promoter sequences. 
For an example with 4 probe genes and $n=5$, Figure 3a represents this step as a graph, where $g1, \dots g4$ are the probe genes (red nodes) and each probe-node is connected by a directed edge to 5 nodes, which are the genes associated with the 5 promoter sequences that are most similar to the probe-node g by the D2z measure. The edges in the graph represent potential co-regulation between the nodes. In this example, the four probe genes produce four clusters, each with 5 arrows from the probe gene to 5 nodes.
The probe genes together with the top $n$ genes for each probe are then used to construct a co-regulation graph as follows. Common nodes between the clusters for the different probe genes are identified (Figure 3b) 
and only nodes that belong to more than one cluster are kept (Figure 3c). In the example, the final graph, Figure 3d, is the co-regulation graph for the probe genes $g1,\dots g4$, where a node in the middle column is colored green if it belongs to at least three clusters and light-blue if it belongs to exactly 2 clusters. 

A highly connected graph means that the probe genes have many common co-regulated genes and therefore these genes are likely to be part of a regulatory network. The example in figure 5 shows that $g1$ and $g3$ are potentially co-regulated with genes A and X (although the set of TFs involved in the co-regulations may be different in each case) and that $g2$ and $g4$ are directly co-regulated. In this example, gene A seems to be commonly co-regulated with most of the genes in the probe gene set and may be a potential hub for the network.
\subsubsection*{Assessing significance}
The significance of the number of commonly co-regulated genes in the co-regulation graph was assessed using co-regulation graph produced from random genes as follows. 
For a co-regulation graph associated with $N$ probes with top $n$ hits per probe, we picked $N$ sets, each of $n$ random genes from DBTSS. Each of these sets corresponds to the top $n$ genes for a probe. We constructed the co-regulation graph as before, using the $N\times n$ genes, together with the $N$ ÒprobeÓ nodes. The p-values were computed using 100,000 such random samples. 

\section*{Acknowledgments}
EJ and MRK were partially supported by the NSF grant DBI-0835718.


\newpage

\section*{Figures}

\begin{figure}
\begin{center}
\includegraphics[width=10.cm,height=10.cm,angle=0,clip]{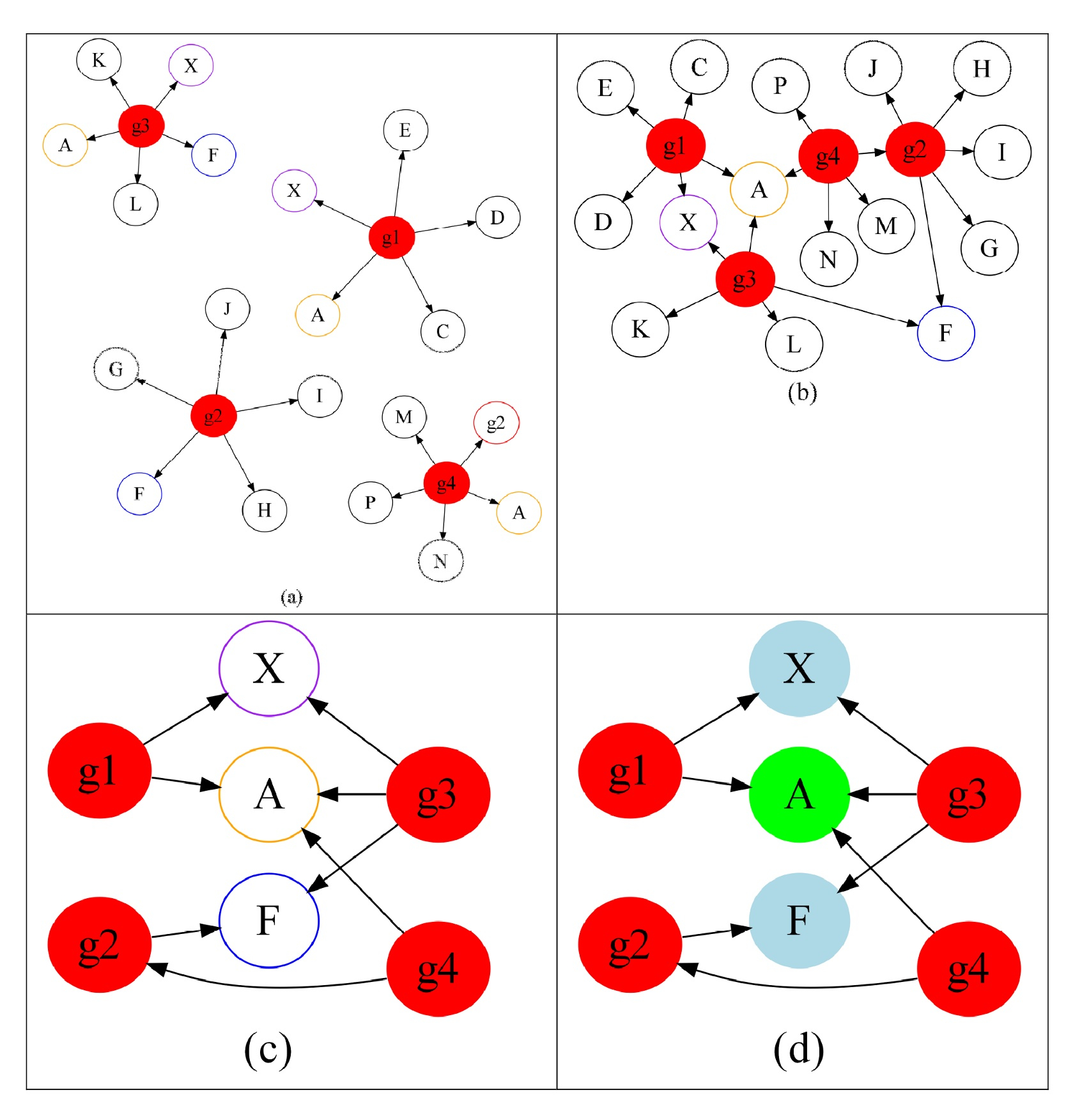}
\caption{Constructing a co-regulation graph.  In this example, the red nodes are the 4 probe genes, $g1, \dots g4$, and the top 5 scoring genes are used for each probe.  The edges in the graph represent potential co-regulation relationship between the nodes. The four probe genes produce four clusters, each with 5 arrows from the probe gene to 5 nodes (a). Common nodes between the clusters for the different probe genes are identified (b) 
and only nodes that belong to more than one cluster (the commonly co-regulated genes) are kept (c). 
The final graph (d) is the co-regulation graph for the probe genes, where the nodes in the middle column are the commonly co-regulated genes. 
The green nodes are genes that are commonly co-regulated by 3 or more probes and the light-blue are commonly co-regulated by 2 probes.
}
\label{fig3}
\end{center}
\end{figure}

\newpage

\section*{Tables} 

\begin{table}[!ht]
\caption{ 
\bf{Genes involved in mental retardation which are commonly co-regulated with the gene set associated with g }}
\begin{tabular}{|c|c|}
\hline
GENE NAME  & ENTREZ GENE ID\\
\hline
AMMECR1	& 9949 \\
\hline
L1CAM	& 3897 \\
\hline
PHF6	& 84295 \\
\hline
ARHGEF6	& 9459 \\
\hline
ALDH3A2  &	224 \\
\hline
ATRX	& 546 \\
\hline
DLG3	& 1741 \\
\hline
GRIK2 &	2898 \\
\hline
HSD17B10	& 3028 \\
\hline
MED12	& 9968 \\
\hline
SYNGAP1	& 8831 \\
\hline
ZEB2	& 9839 \\
\hline
ZNF41 &	7592 \\
\hline
PHF8	& 23133 \\
\hline
PAK3	& 5063 \\
\hline
\end{tabular}
\label{Table1}
 \end{table}

\end{document}